\documentclass[%
 reprint, amsmath, amssymb, aps, superscriptaddress, prx]{revtex4-2}
\usepackage{graphicx}
\usepackage{multirow}
\usepackage{dcolumn}
\usepackage{bm}
\usepackage{hyperref}
\usepackage[mathlines]{lineno}
\usepackage{physics}
\usepackage{float}
\usepackage{diagbox}
\usepackage{booktabs}
\usepackage[normalem]{ulem}
\usepackage{resizegather}
\usepackage{newtxmath}

\hypersetup{
    colorlinks=true,
    linkcolor=red,
    filecolor=magenta,
    citecolor=blue,
    urlcolor=blue,
}

\newcommand{\expect}[1]{{\langle #1 \rangle}}

\newcommand{\hide}[1]{}

\newcommand{\IITB}{Department of Physics, Indian Institute of Technology Bombay, Powai, Mumbai, Maharashtra 400076, India}
\newcommand{\Quicst}{Centre of Excellence in Quantum Information, Computation, Science and Technology,
Indian Institute of Technology Bombay, Powai, Mumbai, Maharashtra 400076, India}

\begin{document}

\title{Quadratic power enhancement in extended Dicke quantum battery}

\author{Harsh Sharma} \affiliation{\IITB}
\author{Himadri Shekhar Dhar} \affiliation{\IITB} \affiliation{\Quicst}


\begin{abstract}
We demonstrate a quadratic enhancement of power in a battery consisting of $N$ two-level systems or spins interacting with two photonic cavity modes, where one of the modes is in the dispersive regime. In contrast to Dicke batteries, the power enhancement arises from a $N^2$ scaling of both quantum correlations and speed of evolution, thus highlighting genuine quantum advantage. Moreover, this hybrid setup is experimentally realizable and ensures that power enhancement is not achieved at significant cost to energy efficiency, while allowing for greater tunability and stable operation in the presence of noise. 
\end{abstract}

\maketitle

\section{Introduction}
Quantum batteries~\cite{Campaioli2018,Campaioli2024,Camposeo2025} are devices that store energy and leverage quantum phenomena such as entanglement~\cite{Alicki2013} and coherence~\cite{Monsel2020} to achieve superior performance over classical setups. 
Over the past decade, considerable research has been done 
on the properties of quantum batteries~\cite{Campaioli2024}, including the
study of maximal energy storage~\cite{Andolina2019} and faster~\cite{Ferraro2018,Rossini2020,Pokhrel2025},  stable~\cite{Rosa2020} and controlled~\cite{Gherardini2020,Mitchison2021} charging protocols.
These storage devices have been considered across a broad range of quantum platforms, including spin chains~\cite{Le2018}, quantum dots~\cite{Wenniger2023}, superconducting qubits~\cite{Santos2019, Dou2023}, and organic semiconductors~\cite{Quach2022}, 
with envisioned applications in quantum communication, sensing, and on-chip energy management~\cite{Chiribella2021,Auffeves2022}.

A key property of a quantum battery, often associated with quantum advantage, is ``power'' i.e., the rate at which the battery can be excited to store energy.
For instance, in quantum batteries consisting of $N$ independent ``cells'', which are typically two-level quantum systems such as spins or atoms, the net power scales as $N$. However, using nonlocal or collective quantum effects the power can often be made superlinear~\cite{Ferraro2018}. For instance, spin models with charging mediated by highly correlated Hamiltonian~\cite{Rossini2020} or dissipative processes~\cite{Pokhrel2025} have demonstrated powers that scale as $N^2$. Recent studies~\cite{JuliaFarre2020} have identified two key pathways to generate high power in quantum batteries -- energy variance in the battery and the speed of evolution. While the former quantifies the role of purely quantum effects such as entanglement to achieve higher power, the latter simply captures the efficiency of the charging mechanism. Very few physically realizable quantum batteries demonstrate power enhancement due to genuine quantum advantage~\cite{JuliaFarre2020, Rosa2020}.

One of the most widely studied models is the Dicke quantum battery~\cite{Ferraro2018,Seidov2024}, motivated in large part by its hybrid architecture and feasibility for implementation in cavity quantum electrodynamics (QED) based experiments~\cite{Raizen1989,Bernardot1992,Brennecke2007,Fink2009,Quach2022}. Based on the $N$ spin Dicke model~\cite{Dicke1954,Hepp1973}, the battery is charged through its collective interaction with light. 
For coherent charging, Dicke batteries exhibit $N^{3/2}$ scaling in power~\cite{Ferraro2018,Seidov2024}, while an enhancement of $N^2$ was recently observed for dissipative charging protocols~\cite{Pokhrel2025}. 
While this is an advantage, both these models rely on speed of evolution governed by the charging, rather than genuine quantum advantage arising from correlations in the battery. Following Ref.~\cite{JuliaFarre2020}, the primary contribution to power for both coherent and dissipative models come from the Fisher information, which scales as $N^2$ and $N^3$, {whereas the variance of the battery Hamiltonian scales only as $N$~\cite{Pokhrel2025} or the power is not optimal~\cite{JuliaFarre2020}.}
This implies that the internal state of the battery behaves  qualitatively similar to $N$ independent systems with no advantage in power arising from quantum correlation. 
{Another notable drawback in power enhanced Dicke quantum batteries is the poor energy efficiency~\cite{Gemme2023,Pokhrel2025}, which occurs due to significant loss in the charging process and leads to substantially less energy being extractable.}

In this work, we present an experimentally realizable quantum battery that achieves $N^2$ power scaling with improved energy efficiency, based on an extended Dicke model. In contrast to other Dicke batteries, both the Fisher information and the energy variance scale quadratically, which shows that the battery depends on quantum correlations to deliver high power, thus providing genuine quantum advantage. Similar to the Dicke model, our battery involves $N$ non-interacting two-level systems or spins. However, the spins now interact with two photonic cavity modes, where one of the modes is in the dispersive regime, while the other is initialized with $N$ photons, as is the case with a conventional Dicke quantum battery~\cite{Ferraro2018}.  
The dispersive cavity introduces strong non-linearity in the spin Hamiltonian and generates quantum correlations in the state of the battery. By controlling the strength of these induced spin-spin correlations, the battery can operate at different regimes to optimize for power and energy, and provide stability in the presence of decoherence or noise in the spins.

The remainder of the paper is organized as follows. In Sec.~\ref{SecII} we introduce the quantum battery model, with physical platforms for implementation  discussed in Sec.~\ref{SecIII}. The charging dynamics and analytical study of power for the extended Dicke battery is described in Sec.~\ref{SecIV}. We discuss our main results in Sec.~\ref{SecV} and end with conclusion in Sec.~\ref{SecVI}. 

\section{The Battery Model\label{SecII}}
Our battery design consists of an ensemble of $N$ identical two-level systems or spins with frequency $\omega_s$, coupled to a two-mode cavity with mode frequencies $\omega_b$ and $\omega_c$ and coupling strength $g_b$ and $g_c$.
The first cavity is far-detuned from the spins (i.e., $|\omega_b -\omega_s|\gg g_b$), with anisotropy in the counter-rotating (excitation non-conserving) terms.
The Hamiltonian of the system is given by ($\hbar = 1$)
\begin{multline}
    \hat{H} = \omega_c \hat{c}^\dagger \hat{c} + \omega_s \hat{J}_z + 2 g_c (\hat{c}^\dagger + \hat{c})\hat{J}_x \\+ \omega_b \hat{b}^\dagger \hat{b} + g_b\left( \hat{b} \hat{J}_{+} + \hat{b}^\dagger \hat{J}_{-} + \delta_b \hat{b} \hat{J}_{-} + \delta_b \hat{b}^\dagger \hat{J}_{+} \right),
    \label{Eq:Hlab}
\end{multline}
where $\hat{J}_\alpha = \sum_{i=1}^N \hat{\sigma}_{i,\alpha}/2$ are the collective spin operators, and $\hat{\sigma}_{i,\alpha}$, for $\alpha \in \{x,y,z,+,- \}$, are the Pauli spin-$1/2$ operators and their linear combinations, acting on the $i^{\text{th}}$ spin. 
{$\hat{b}$ and $\hat{c}$ are the annihilation operator of the cavity modes $b$ and $c$, respectively.} Moreover, $|\delta_b|\leq1$ is the anisotropy parameter.
Since generating anisotropy generally requires external driving fields (for e.g. see Appendix~\ref{AppA}), it is convenient to replace cavity and qubit frequencies with their corresponding detunings from an external drive with frequency $\omega_d$, i.e, $\omega_{b/c/s}\rightarrow \Delta_{b/c/s}$.
In absence of any decoherences, the total angular momentum $\hat{J}^2 = \hat{J}_x^2 + \hat{J}_y^2 + \hat{J}_z^2$ in the system is conserved.
{As the initial state of the battery is the completely unexcited spin state, which lies in the symmetric Dicke subspace with $J=N/2$, the state of the battery during charging and discharging also remains in this subspace.}

The battery-charger design constitutes a hybrid quantum system~\cite{Kurizki2015,Clerk2020}, with the spins ensemble constituting the primary battery, governed by the Hamiltonian $\hat{H}_B = \omega_s \hat{J}_z$, while the interaction Hamiltonian describes the charging mechanism. 
The two cavity modes play specific roles -- 
mode $c$ is used to coherently transfer energy to the spins, while mode $b$ mediates the spin-spin interactions in the battery.
{The latter is evident after Schrieffer-Wolff transformations~\cite{Schrieffer1966,Bravyi2011}, using which mode $b$ can be adiabatically eliminated in the dispersive regime ($|\Delta_b -\Delta_s|\gg g_b\sqrt{N}$).}
Such transformations have been previously studied in the context of Dicke batteries~\cite{Groszkowski2020,Gemme2023}. Refer to Appendix~\ref{AppA1} for more detailed calculations. 
The final transformed Hamiltonian with a single cavity mode coupled to the spin ensemble, with two-axis-twist like spin squeezing, is given by
\begin{equation}
    \hat{H}_{sq} \approx \Delta_c \hat{c}^\dagger \hat{c} +\tilde{\Delta}_s \hat{J}_z + 2g_c (\hat{c}^\dagger + \hat{c})\hat{J}_x
    -\chi_b \left(f_{\delta_b}^+\hat{J}_x^2 + f_{\delta_b}^-\hat{J}_y^2 \right),
    \label{Eq:HamTAT}
\end{equation}
where $\tilde{\Delta}_s = \Delta_s - \chi_b(1-\delta_b^2\Delta_-/\Delta_+)$, $\chi_b = {g_b^2/\Delta_-}$, $\Delta_\pm = \Delta_b \pm \Delta_s$ and $f_{\delta_b}^\pm = 1\pm\delta_b+(\delta_b^2\pm\delta_b)\Delta_-/\Delta_+$. It can be seen that the widely studied Dicke quantum battery~\cite{Ferraro2018} is recovered in the limit $\chi_b/g_c \rightarrow 0$ or when the coupling between mode $b$ and spins in the battery is turned off i.e., $g_b=0$ in Eq.~\eqref{Eq:Hlab}. Also, the anisotropy parameter $\delta_b$ helps us control the direction of spin-spin interactions, which is proportional to $\hat{J}_x^2, \hat{J}_y^2, \hat{J}_z^2$ for $\delta_b=1,-1,0$, respectively.

{Note that we do not scale the coupling strength $g_c$ and $g_b$ by $1/\sqrt{N}$ in our model, which is often termed the ``thermodynamic limit''~\cite{JuliaFarre2020,Rosa2020}. Experimental studies have clearly exhibited $\sqrt{N}$ scaling of interaction in solid state systems~\cite{Kubo2010,Schuster2010} and achieved superabsorption based fast charging~\cite{Quach2022}, for very large $N$. 
Recently, it was emphatically argued that for all physically realizable battery models, power can only scale linearly in the infinite thermodynamic limit~\cite{Pokhrel2025}. As such, our focus here is on the experimentally relevant large, but finite $N$ limit. }

\begin{figure}
    \centering
    \includegraphics[width=\linewidth]{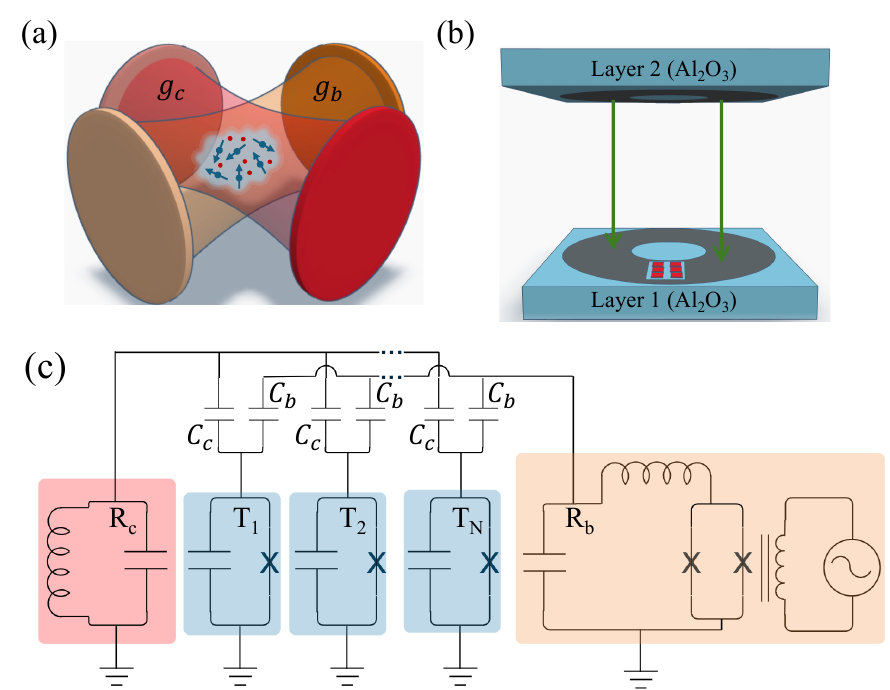}
    \caption{Two mode Dicke quantum battery. (a) Schematic of the quantum battery model containing two cavity modes (red and orange) and the spin ensemble is spatially placed inside the cavity such that it is coupled with both the modes. (b) Whispering-gallery-mode resonator~\cite{Minev2013}, consisting two superconducting ring resonator (black), one placed on each layer (blue), which are seperated by a thin vacuum gap. The qubits are carved directly on the resonator of one layer (red) and out-of-plane fields (green) couple the two layers. (c) Circuit-QED representation of the model. The resonator in red box represents cavity mode $\hat{c}$ and the one in orange box is mode $\hat{b}$, which is parametrically driven through a flux-pumped superconducting quantum interference device (SQUID)~\cite{Yamamoto2008} to get the required anisotropy in mode $b$. The two resonators are coupled to transmon qubits (in blue box) via coupling capacitors $C_{c}$ and $C_{b}$, respectively.}
    \label{fig:1}
\end{figure}

\section{Physical implementation\label{SecIII}}
Implementing our quantum battery requires two key elements (i) a cavity supporting two distinct modes and (ii) a suitable mechanism to generate anisotropy. Multimode cavities are available across several platforms, including ion traps~\cite{Jia2022}, atomic ensembles in optical resonators~\cite{Wickenbrock2013}, and circuit QED~\cite{Minev2016,Chakram2021}, however integrating them with anisotropic interaction is challenging. One way to do it is via two-photon parametric drive, which can generate anisotropy with $\delta_b=-\tanh r$, where $r$ is the squeezing parameter related to amplitude of drive and the detuning between the cavity and drive frequency $\Delta_b$ (see Appendix~\ref{AppA}).
Recent experiments have realized such drives in ion traps~\cite{Burd2021,Affolter2023,Burd2024} and superconducting circuits~\cite{Villiers2024}, achieving maximum squeezing of $r \approx 1.4$ and $r \approx 0.8$, respectively. 
Given that circuit QED has been used for preliminary quantum battery experiments~\cite{Hu2022}, and is also suitable for implementing a two-mode cavity and parametric drive, we focus on a circuit QED implementation of our protocol.

Figure~\ref{fig:1}(a) shows the visualization of the two-mode quantum battery. The model can be potentially realized in planar multilayer circuit QED architectures~\cite{Minev2016,Brecht2016}, for example, using whispering-gallery mode resonators~\cite{Minev2013} (see Fig.~\ref{fig:1}(b)). These are ring transmission line resonators, etched on thin vacuum separated substrate surface layers. These ring resonators lack circular symmetry and their coupling with the qubits can be tuned depending on the position of aperture where qubits (shown in red) are carved~\cite{Minev2016}. In these systems, spin and resonator frequencies are in the gigahertz range, with couplings from a few to several hundred megahertz and decay rates on the order of a few hundred kilohertz~\cite{Minev2016}. 
Figure~\ref{fig:1}(c) sketches a circuit QED realization of the system, where a transmon qubit (the “spin”) is capacitively coupled to a microwave resonator (the “charger”) supporting two modes, $b$ and $c$. Mode $c$ provides the primary light–matter coupling for charging, while mode $b$ is likewise coupled to the transmon and incorporates a flux-pumped SQUID that implements the required two-photon (parametric) drive~\cite{Yamamoto2008,Leroux2018}. In addition to parametric driving, anisotropy can also be generated via periodic frequency modulation, where two periodic drive fields are used to generate tunable anisotropic interactions. This allows a greater control on the system's initial parameters just by adjusting the amplitude, frequency and phase of driving fields~\cite{Wang2019}. 

\section{Charging Dynamics\label{SecIV}} 
Initially, the quantum battery is completely discharged (all spins are unexcited), while the charging cavity mode $c$ has $N$ photons and $b$ is in vacuum state. The state at $t=0$ can be written as $\ket{\Psi(0)} = \ket{N/2,-N/2}_{s} \otimes \ket{N}_{c}\ket{0}_{b}$, where the discharged battery is written in the collective angular momentum basis with $J=N/2$ and $M=-N/2$ and the cavity modes in the Fock state basis.
As such, all the energy is initially stored in the cavity modes, which serve as the charging unit. At $t=0$, the energy in the charger is $E_{T} = N\omega_c$, where $N\omega_c$ is the energy in mode $c$ that scales as number of spins $N$ in the battery.
In the transformed frame, the dynamics of the battery is governed by the Hamiltonian $\hat{H}_{sq}$ in Eq.~\eqref{Eq:HamTAT}, which is a good approximation of our system
as long as $g_b\sqrt{N} \ll |\Delta_-|$. Appendix~\ref{AppA2} compares the dynamics of the approximate model with the original Hamiltonian in Eq.~\eqref{Eq:Hlab}.

After charging time $t_c$, the system evolves to state $\ket{\Psi(t_c)} = \text{Exp}\left[-i \hat{H}_{sq} t_c\right] \ket{\Psi'(0)}$, where $\ket{\Psi'(0)}$ is the corresponding initial state in the transformed frame. The energy {transferred to the spins in the battery during} this time is given by $E_B(t_c) = \omega_s \left( \expect{\hat{J}_z(t_c)} - \expect{\hat{J}_z(0)} \right)$.
{Using the energy, we can calculate the key figure of merit related to our battery, i.e., the average charging power. It is defined as $\bar{P}(t_c) = E_B(t_c)/t_c$. 
In the Heisenberg picture, the (instantaneous) power ${P}$ is given by
\begin{align}
    P(t_c) &\propto {d \over dt}\expect{\hat{J}_z} = i\expect{[\hat{H}_{sq},\hat{J}_z]} \nonumber\\
    &= 2g_c\expect{(\hat{c}^\dagger + \hat{c})\hat{J}_y}-2\chi_b\delta_b \left(1+\frac{\Delta_-}{\Delta_+}\right)\expect{\{\hat{J}_x,\hat{J}_y\}},
    \label{Eq:Pwr}
\end{align}
where $\{x,y\}=x.y+y.x$ is the anti-commutator. 
It is evident from the above expression that spin-spin correlations have a positive effect on power when $\delta_b=-1$, power is unaffected by it when $\delta_b=0$ and there is a negative effect when $\delta_b = 1$.

Using the fact that expectation values of the operators are upper bounded by the operator norms, i.e., $\norm{\hat{J}_{x,y}} \sim \mathcal{O}(N)$, $\norm{\{\hat{J}_{x}, \hat{J}_{y}\}} \sim \mathcal{O}(N^2)$ and for our initial state $\norm{\hat{c}^\dagger + \hat{c}} \sim \mathcal{O}(\sqrt{N})$, the charging power can be upper bounded as,
\begin{equation}
    |P(t)| \lesssim a_{c} g_c N\sqrt{N} + a_{b} \chi_b N^2,
    \label{Eq:PwrBound}
\end{equation}
for some system parameter dependent constants $a_{c}$ and $a_{_b}$. This tells us that for the charging Hamiltonian in Eq.~\eqref{Eq:HamTAT}, {the quantum battery can be charged superlinearly, as high as a quadratic $N^2$ scaling.} 
Irrespective of the charging Hamiltonian, the power of a quantum battery arises from two key quantities~\cite{JuliaFarre2020} -- the variance of the battery Hamiltonian ${\Delta \hat{H}_B^2}$ and the Fisher information $I_E$, and it follows the bound $\expect{P}_\tau \leq \sqrt{\expect{\Delta \hat{H}_B^2}_\tau \expect{I_E}_\tau}$, where $\expect{\hat{O}}_\tau \equiv \tau^{-1} \int_0^\tau \hat{O}(t) dt$ is the time average of the observable $\hat{O}$. Importantly, $\Delta \hat{H}_B^2$ captures the quantum correlations in the battery state and represents contribution to power arising from genuine quantum advantage. On the other hand $I_E$ is the Fisher information, and captures the speed at which energy flows through the system. We also study the saturation of power bound using the quantity $P_{\text{sat}} = \expect{P}_\tau/\sqrt{\expect{\Delta \hat{H}_B^2}_\tau \expect{I_E}_\tau}$ and define bound saturation $\beta_b$ as the scaling of $P_{\text{sat}}$ with $N$ i.e., $P_{\text{sat}}\sim N^{\beta_b}$. $\beta_b < 0$ indicates that the charging power moves further away from the bound as $N$ is increased. It was observed numerically that in the case of Dicke charging model, $\beta_b \approx -0.3$~\cite{JuliaFarre2020}, and this is the main reason for the model not showing genuine quantum advantage in strong coupling regime even though the bound predicts it. 
\begin{figure}[t!]
    \centering    
    \includegraphics[width=\linewidth]{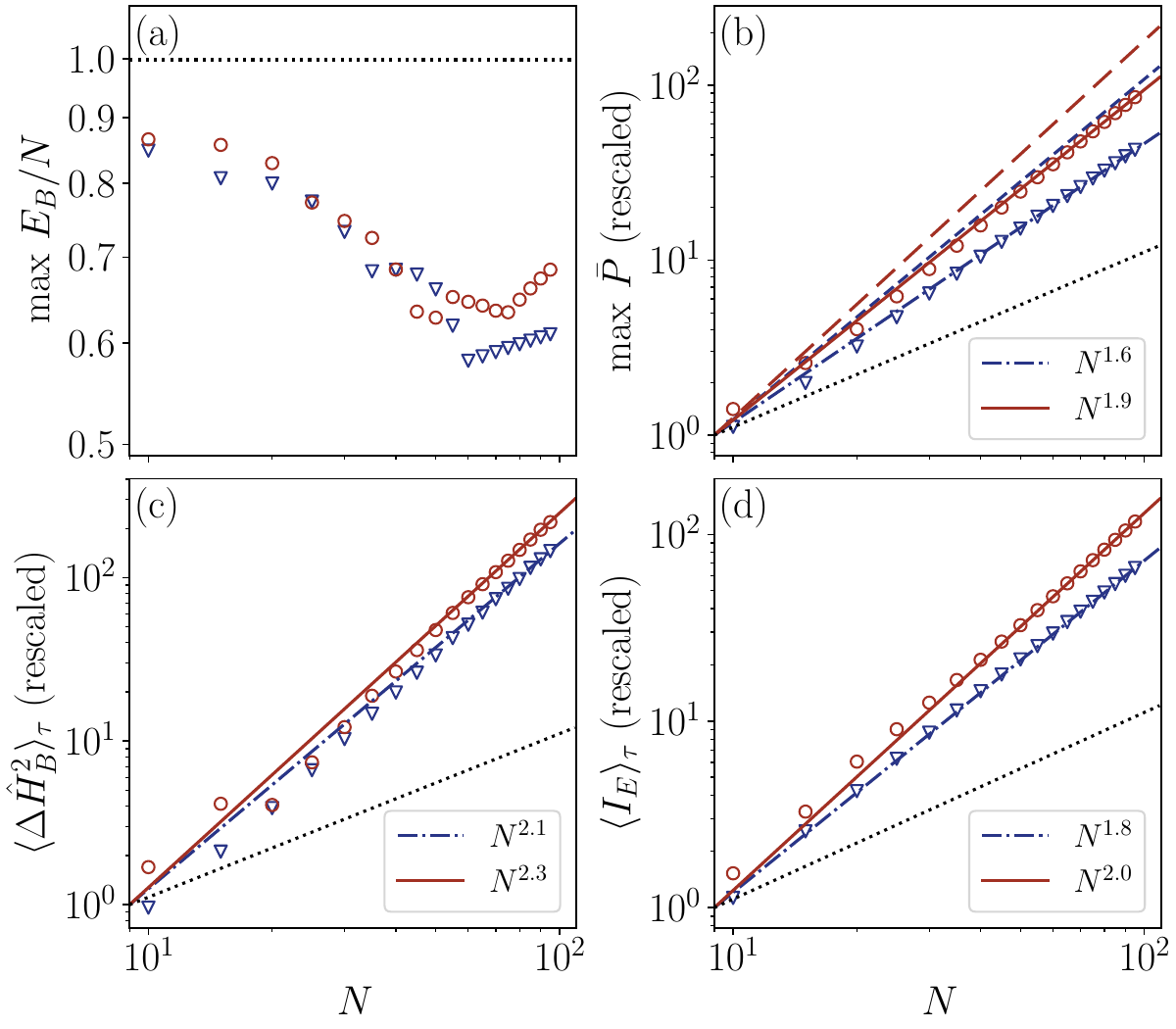}
    \caption{Dynamics of two-mode QB. (a) Maximum charging energy per spin as a function of number of spins $N$ in the ensemble for $\tilde{\Delta}_s=\Delta_c = 0.7~\omega_c$, $\Delta_b= 101~\omega_c$, $g_c=0.05~\omega_c$, $\delta_b=-1$, and $g_b = 0~\omega_c$ (blue triangles) and $g_b=0.9~\omega_c$ (red circles). (b) Maximum charging power as a function of $N$ for same parameters, with short and long dashed lines representing the upper bound on charging power~\cite{JuliaFarre2020}, for $g_b/\omega_c = 0, 0.9$, respectively. (c) and (d) Average battery variance $\expect{\Delta \hat{H}_B^2}_\tau$ and Fisher information $\expect{I_E}_\tau$, respectively, computed at time $\tau$ where power is maximum. Legends in (b)-(d) show the scaling of the quantities with $N$ obtained from best fit curve, with the lines joining the markers being the best fit lines. Black dotted lines in (a-d) represents parallel charging case, where each spin is charged individually.}
    \label{fig2}
\end{figure}

\section{Main Results\label{SecV}}
Figure~\ref{fig2}(a-d) shows the variation of the four quantities described above, i.e., $\max{E_B}$, $\max{\overline{P}}$, $\expect{\Delta \hat{H}_B^2}_\tau$ and $\expect{I_E}_\tau$ with the number of spins in the two-mode quantum battery. This is compared with two typical cases: parallel and collective charging of a Dicke battery.
Figure~\ref{fig2}(a) shows the maximum charging energy in all three cases. The parallel charging is most energy efficient as there is no power enhancement. However, the two-mode battery model performs better than the typical collectively charged Dicke battery for large $N$ as it provides similar energy but with higher power, and is therefore more efficient.
In addition to this, we also see additional enhancement in charging power in comparison to the Dicke battery (see Fig.~\ref{fig2}(b)). 
This happens mostly due to presence of spin-spin correlations in the two-mode battery, which helps it to better saturate the bound in Ref.~\cite{JuliaFarre2020}. Some enhancement in power also comes from faster charging, which is evident from slightly higher scaling of Fisher information in Fig.~\ref{fig2}(d).
\begin{figure}[t]
    \centering
    \includegraphics[width=\linewidth]{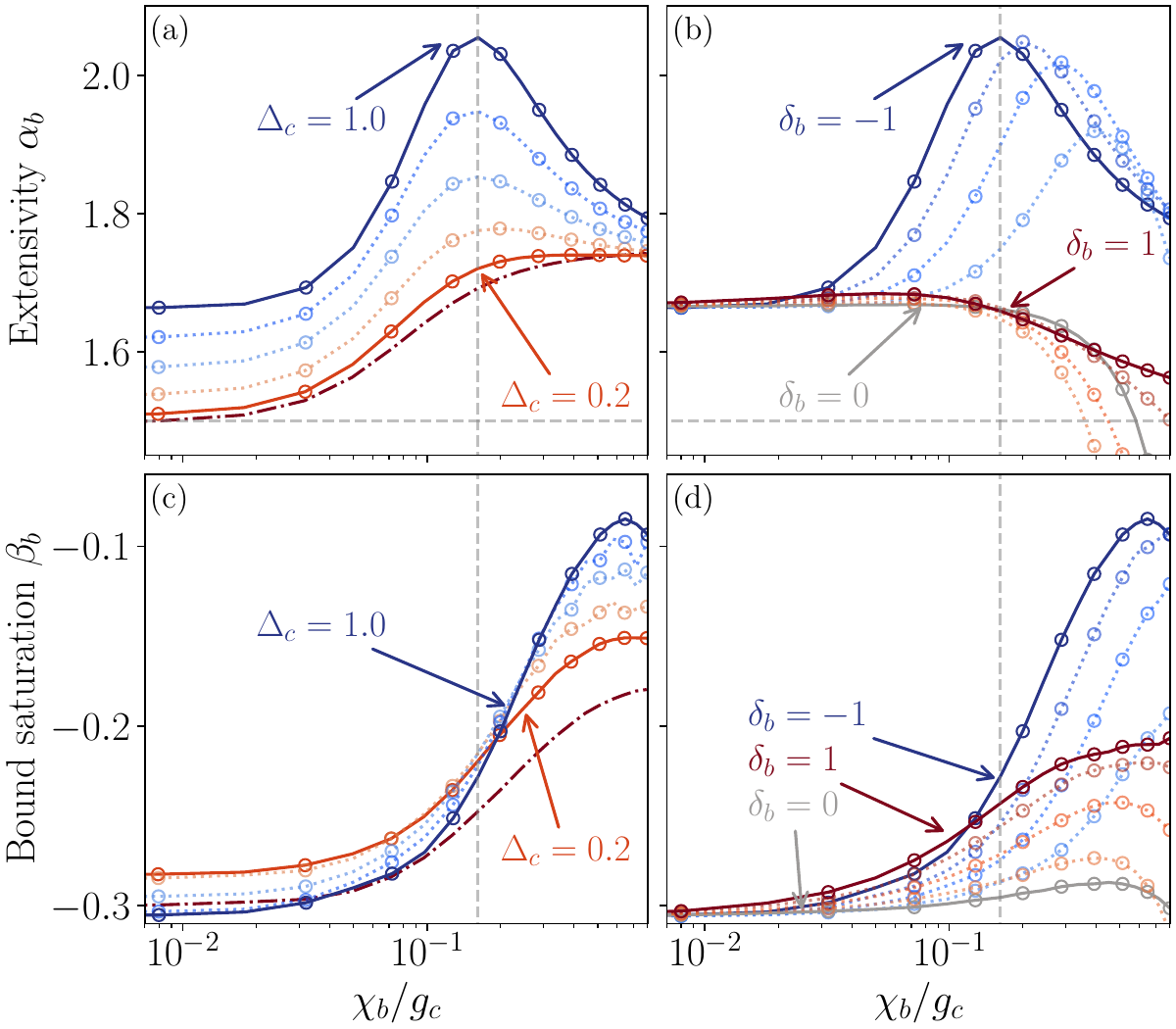}
    \caption{Scaling of bounds in the two-mode quantum battery. (a) Scaling of power (or extensivity) of the QB as a function of relative strength of spin-spin interaction $\chi_{b}$. Different curves corresponds to different $\tilde{\Delta}_s = \Delta_c$ for $\delta_b=-1$, equally distributed between ${\Delta}_c/\omega_c = 0.2$ and $1$. Strength of spin-spin interaction is controlled using $g_b/\omega_c$, which varies from $0.2$ to $2$, with $g_b/\omega_c = 0.9$ marked by vertical dashed line. (b) Same as (a) but for $\Delta_c/\omega_c=1.0$ and different anisotropy $\delta_b$, equally distributed between $-1$ and $1$. (c)-(d) Scaling of $P_{\text{sat}}$ (or bound saturation) as a function of spin-spin interaction strength, for same parameters as in (a) and (b), respectively.
    Other simulation parameters are $\Delta_b = 101~\omega_c$ and $g_c = 0.05 \omega_c$ and $\alpha_{r_b}$ is estimated using data from $N = 75$ to $100$ spins. Dot-dashed red curve in (a) and (c) marks the minimum enhancement achievable in the two mode case.}
    \label{fig3}
\end{figure}

Figure~\ref{fig3}(a) shows how $\alpha_{b}$ ($P \sim N^{\alpha_b}$) varies with relative spin-spin interaction strength $\chi_{b}/g_{c}$. Interestingly, we find that when both the strengths are weak {(i.e., $\chi_{b},g_{c}\ll \omega_c$)}, 
the power scales as $\mathcal{O}(N^2)$ for some optimal value of mode $b$ coupling $g_b$. However, at same $\chi_b$, increasing the relative Dicke coupling $g_c/\Delta_c$ of mode $c$ generally decreases the maximum value of $\alpha_{b}$. This is because overcoming Dicke interaction then requires relatively larger spin-spin interaction strength. Figure~\ref{fig3}(c) shows the bound saturation defined in previous section as a function of spin-spin interaction strength. It increases (gets closer to zero) as $\chi_b$ is increased, indicating {that power} {is closer to} the theoretical bound at large $N$, thereby exhibiting genuine quantum advantage. Both quantities are analyzed for different anisotropy $\delta_b$ in Fig.~\ref{fig3}(b) and (d). As expected from Eq.~\eqref{Eq:Pwr}, even though bound saturation $\beta_b$ is better when $|\delta_b|>0$, power enhancement is observed only when $\delta_b<0$.

Now, Eq.~\eqref{Eq:PwrBound} suggests that the power should scale quadratically for all high values of spin-spin interaction strength $\chi_b$. However, it is observed that quadratic power scaling is only achieved for some optimal parameters, as seen in Fig.~\ref{fig3} and it decreases with spin-spin interaction strength when $\chi_b/g_c \gtrsim 1$. To qualitatively understand this, we analyze the Hamiltonian in Eq.~\eqref{Eq:HamTAT} under Holstein-Primakoff approximation~\cite{Holstein1940}, where the collective spin operators are approximated as Bose operators obeying $\left[\hat{s},\hat{s}^\dagger \right] = 1$.
For large $J$ or when bosonic excitations in the system is small i.e., $\expect{\hat{n}_s}=\expect{\hat{s}^\dagger s}\ll J$, we can approximate $\hat{J}_\pm \approx \sqrt{2J} \hat{s}^{\dagger/}$, where $\hat{J}_\pm = \hat{J}_x\pm i \hat{J}_y$, and redefine the Hamiltonian in Eq.~\eqref{Eq:HamTAT} as a system of two coupled oscillators. Under the spin Bogoliubov transformation,  $\hat{s}' = \cosh{r_s} \hat{s} - \sinh{r_s} \hat{s}^\dagger$, with $\tanh{2r_s} = \kappa_b/\tilde{\Delta}'_s$ and with rotating wave approximation, we have
\begin{equation}
    \hat{H}_{\text{eff}} \approx \Delta_c \hat{c}^\dagger \hat{c} + \Delta'_s \hat{s}'^\dagger s' + \Omega'_c (\hat{c}\hat{s}'^\dagger + \hat{c}^\dagger \hat{s}'),
    \label{Eq:HamBS}
\end{equation}
where $\Delta'^2_s = (\tilde{\Delta}_s - N\chi_b\delta_b)^2 - N^2\chi_b^2$, $\Omega'_c = g_c \sqrt{N}e^{r_s}$, $r_s = \ln{\left( (\tilde{\Delta}_s-N\chi_b(\delta_b-1))/(\tilde{\Delta}_s+N\chi_b(\delta_b+1))\right)^{1/4}}$ and we hve assumed $\Delta_-\ll\Delta_+$. Surprisingly, this is exactly the equation of two modes coupled via a beam-splitter interaction and with an exponentially amplified coupling strength. Though this solution is strictly valid for low excitations, it gives us important insight about the dynamics of the system. Note that spin-spin correlations (exponentially) enhances the coupling between the cavity and spin-boson mode, it also increases detuning $\Delta = |\Delta_c - \Delta'_s|$ between the two. While the former enhances the power by decreasing the charging time, the latter has opposite effect as it hinders efficient energy transfer between the two modes. As a result, though $\alpha_b$ initially increases on increasing spin-spin interaction strength, eventually the detuning effects dominates and $\alpha_b$ starts decreasing.

\subsection{Effects of dissipation\label{SecVA}}
In general both the spins and the cavity in a hybrid system are subject to intrinsic losses, i.e., they naturally lose coherence over time. Consequently, the dynamics of such a system are most appropriately described by a Lindblad master equation~\cite{Breuer2007}  
\begin{equation}
    \frac{d\hat{\rho}}{dt} = -i[\hat{H}_{sq},\hat{\rho}] 
    + \kappa_c \mathcal{L}_{\hat{c}}[\hat{\rho}]
    + \Gamma_b \mathcal{L}_{\hat{J}_b}[\hat{\rho}]
    + \sum_{j=0}^1 \sum_{k=1}^N \gamma_{j} \mathcal{L}_{\hat{x}^j_k}[\hat{\rho}],
    \label{Eq:lind}
\end{equation}
where $\hat{\rho}$ denotes the density matrix of the system, and the Lindblad superoperators are defined as $\mathcal{L}_{\hat{x}}[\hat{\rho}] = \hat{x}\hat{\rho} \hat{x}^\dagger - \tfrac{1}{2}\{\hat{x}^\dagger \hat{x}, \hat{\rho}\}$, with operators $\hat{x}^0_k = \hat{\sigma}^z_k$ and $\hat{x}^i_k = \hat{\sigma}^-_k$ for the $k^{th}$ spin.
Here, $\kappa_{c/b}$ is the photon loss rate from the cavity mode $c/b$, $\gamma_0$ is the pure dephasing rate leading to coherence loss without population transfer, and $\gamma_{1}$  is the radiative decay rate characterizing energy relaxation.
In addition to these, cavity mode $b$ also induces collective dissipation on spins given by $\hat{J}_b = \hat{J}_- - \delta_b (\Delta_-/\Delta_+)\hat{J}_+$, which occurs with rate $\Gamma_b = \kappa_b \chi_b/\Delta_-$~\footnote{Cavity mode $b$ induced dissipation can be found by studying the behavior of $e^{\hat{S}}\hat{b}e^{-\hat{S}}$ under Lindblad dynamics, where $\hat{S}$ is the generator of Schrieffer-Wolff transformation. Similar analysis can be found in supplemental material of Ref.~\cite{Groszkowski2020}.}.
These dissipation channels are unavoidable in realistic experimental implementations and therefore must be incorporated into any accurate description of hybrid spin-cavity dynamics. 
For decoherence rates typical of whispering-gallery-mode resonators~\cite{Minev2016}, we find that the charging dynamics is essentially unaffected up to the first peak in the stored energy. Since our analysis is restricted to this short-time regime, our results should therefore remain valid for a physical realization of the battery in such systems.
\begin{figure}
    \centering
    \includegraphics[width=0.9\linewidth]{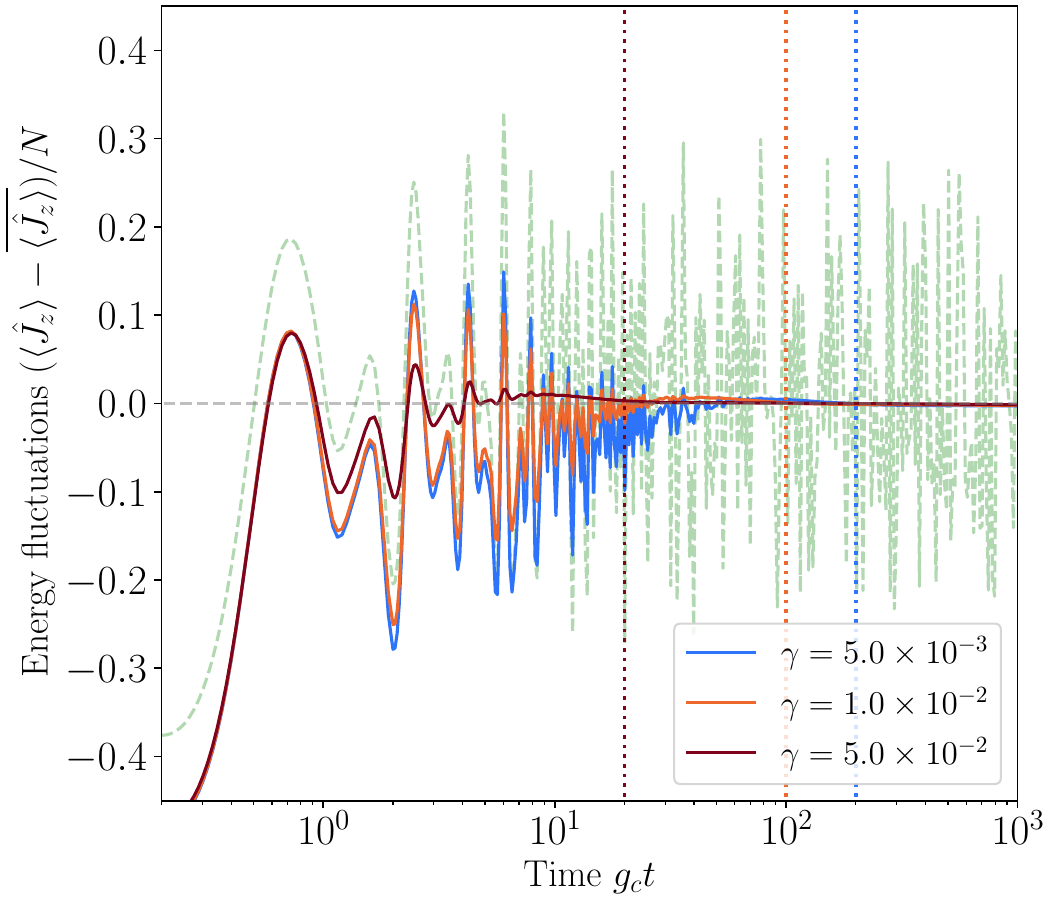}
    \caption{Decoherence dynamics in the two mode quantum battery. Battery energy fluctuations $\expect{\hat{J}_z}-\overline{\expect{\hat{J}_z}}$ as a function of time for $\chi_b = 0.075~\omega_c$ with (solid) and without (dashed) decoherence. The dynamics was simulated for $N=10$, $\Delta_b=100$, $\delta_b\approx-1$, $g_c = 0.05$, $\kappa_{c/b} = 10^{-5}$ and $\gamma_{-1}=2\gamma_{0}=\gamma$ (in units of $\omega_c=1$). Vertical dotted lines marks the approximate time after which battery energy stabilizes.}
    \label{fig4}
\end{figure}

Now, we deliberately consider relatively large single-spin error rates compared to the cavity decay to highlight their impact on the battery dynamics. Figure~\ref{fig4} shows the battery energy fluctuations as a function of the charging time in the presence of such decoherences.  
When the dynamics is close to collective charging of a Dicke battery, the battery energy decreases at long charging times, which manifests as negative fluctuations (not shown here). But for the two-mode quantum battery, when $\chi_b > g_c$ the battery energy is stabilized and the fluctuations vanish at long times. Moreover, the battery subject to higher decoherence rates reaches this stabilized regime faster. This suggests that appropriately engineered dissipation in such quantum batteries can assist in realizing fast and stable charging and storage under realistic conditions.

\section{Conclusion\label{SecVI}}
Our analysis demonstrates that the two-mode quantum battery offers a clear advantage over conventional Dicke-based architectures. The inclusion of a detuned and anisotropic second mode $b$ enables enhanced collective charging, with the power-scaling exponent $\alpha_{b}$ exhibiting superlinear growth as a function of the spin-cavity coupling $g_b$. This performance gain arises from an improved saturation of the fundamental power bound, facilitated by effective spin–spin interactions in the two-mode setting, a feature absent in the standard DQB. Beyond faster charging, the two-mode model also supports a larger energy storage capacity, establishing it as a versatile platform for achieving tunable superlinear scaling. Importantly, we find that these advantages persist in the presence of decoherences, and fast and stable battery charging can be achieved via engineered dissipation. Taken together, our results highlight the potential of engineered multimode and driven interactions for the practical realization of fast, robust, and high-capacity quantum batteries.

\section*{Acknowledgements}
H.S. acknowledges financial support from the Prime Minister’s Research Fellowship (ID: 1302055), Govt. of India. H.S.D. acknowledges support from SERB-DST, India under a Core-Research Grant (No: CRG/2021/008918) and from IRCC, IIT Bombay (No: RD/0521-IRCCSH0-001).

\appendix
\section{Parametrically driven two mode quantum battery model\label{AppA}}
Consider the similar model as described in the main text, but instead of being anisotropic, the cavity mode $b$ is far-detuned from the spins (i.e., $|\omega_b -\omega_s|\gg g_b$) and is subject to a parametric two-photon drive with frequency $2\omega_d$ and strength $\eta$. Then, the Hamiltonian of this system in the lab frame is given by ($\hbar = 1$)
\begin{multline}
    \hat{H}_{{lab}} = \omega_c \hat{c}^\dagger \hat{c} + \omega_s \hat{J}_z + 2 g_c (\hat{c}^\dagger + \hat{c})\hat{J}_x \\+ \omega_b \hat{b}^\dagger \hat{b} + \left(g_b \hat{b} \hat{J}_{+} + {\eta \over 2} e^{i 2 \omega_d t} \hat{b}^2 + \text{h.c.} \right).
    \label{Eq:HPDr}
\end{multline}
In the frame rotating with frequency $\omega_d$, the Hamiltonian ca be rewritten in {interaction picture} as
\begin{multline}
    \hat{H}_{rot} = \Delta_c \hat{c}^\dagger \hat{c} +\Delta_s \hat{J}_z + 2g_c (\hat{c}^\dagger + \hat{c})\hat{J}_x\\
    + \Delta_b \hat{b}^\dagger \hat{b} + \left(g_b \hat{b}\hat{J}_{+} + \frac{\eta}{2} \hat{b}^2 + \text{h.c.} \right),
    \label{Eq:HamRot}
\end{multline}
where $\Delta_{c,s,b} = \omega_{c,s,b} - \omega_d$. The non-linear terms in the Hamiltonian can be removed by going to the ``squeezed frame" using the squeezing unitary $\hat{U}_{sq} = \exp{-r_b (\hat{b}^2 - (\hat{b}^\dagger)^2)/2}$,
\begin{align}
    \hat{H'}_{sq} &= \hat{U}_{sq} \hat{H}_{rot} \hat{U}_{sq}^{\dagger} \nonumber \\
    &= \Delta_c \hat{c}^\dagger \hat{c} +\Delta_s \hat{J}_z + 2g_c (\hat{c}^\dagger + \hat{c})\hat{J}_x \nonumber\\
    &\quad+ \frac{\Delta_b}{\cosh{2r_b}} \hat{b}_s^\dagger \hat{b}_s + g_b \left( \hat{b}_s \hat{J}_{+}^s+ \text{h.c.} \right),
    \label{Eq:ApHamsq}
\end{align}
where $\tanh{2r_b} = \eta/\Delta_b$, $\hat{b}_s = \cosh{r_b} \hat{b} + \sinh{r_b} \hat{b}^\dagger$ and $\hat{J}_{+}^s = \cosh{r_b} \hat{J}_{+} - \sinh{r_b} \hat{J}_{-}$. This is exactly the Hamiltonian in Eq.~\eqref{Eq:Hlab} of main text, with $\omega_{c/s}\rightarrow \Delta_{c/s}$, $\omega_b \rightarrow \Delta_b \cosh^{-1}{2r_b}$, $g_b \rightarrow g_b \cosh{r_b}$ and $\delta_b=-\tanh{r_b}$.

\section{Cavity induced spin-spin interactions\label{AppA1}}
Consider the Hamiltonian in Eq.~\eqref{Eq:Hlab} of the main text. In the dispersive regime, i.e., when $g_b \sqrt{N} \ll \Delta_-$, we can eliminate the cavity mode $b$ using Schrieffer-Wolff transformation~\cite{Schrieffer1966,Bravyi2011} with (anti-hermitian) generator $\hat{S}=-\hat{S}^\dagger$, such that it block diagonalizes the Hamiltonian in mode $b$. Under such transformation, the Hamiltonian is given by,
\begin{align}
    \hat{H}_{sq} &= e^{\hat{S}} \hat{H} e^{-\hat{S}} \nonumber\\
    &\simeq  \Delta_c \hat{c}^\dagger \hat{c} + \tilde{\Delta}_s \hat{J}_z + 2g_c (\hat{c}^\dagger + \hat{c})\hat{J}_x - \chi_b \left(f_{\delta_b}^+\hat{J}_x^2 + f_{\delta_b}^-\hat{J}_y^2\right) \nonumber\\
    & \quad + {\Delta_b} \hat{b}^\dagger \hat{b} - \chi_b 2(1-\delta_b^2) \hat{b}^\dagger \hat{b} \hat{J}_z \nonumber\\
    & \quad - \chi_b \delta_b \left(1-\frac{\Delta_-}{\Delta_+}\right)\left(\hat{b}^{\dagger^2}+\hat{b}^2 \right)\hat{J}_z,
    \label{Ap:HamSW}
\end{align}
where $\hat{S} = \frac{g_b}{\Delta_-} \left( \hat{b}^\dagger \hat{J}_{-} - \text{h.c.}\right) + \frac{g_b\delta_b}{\Delta_+} \left( \hat{b}^\dagger \hat{J}_{+} - \text{h.c.}\right)$, $\chi_b=g_b^2/\Delta_-$, and only leading order terms in $g_b/(\Delta_b-\Delta_s)$ were retained in the final expression. Note that the last term in a closed system evolution is very small (for $\Delta_s \ll \Delta_b$), which can be safely neglected. The dispersive interaction term $\hat{b}^\dagger \hat{b} \hat{J}_z$ in the Hamiltonian above can be eliminated by choosing a specific spin detuning $\Delta_s$~\cite{Groszkowski2020} or through a suitable dynamical decoupling protocol. After such eliminations, the final Hamiltonian (up to a constant energy term $\hat{b}^\dagger \hat{b}$) takes the form given in Eq.~\eqref{Eq:HamTAT} in the main text.
\begin{figure}[h]
    \centering
    \includegraphics[width=0.9\linewidth]{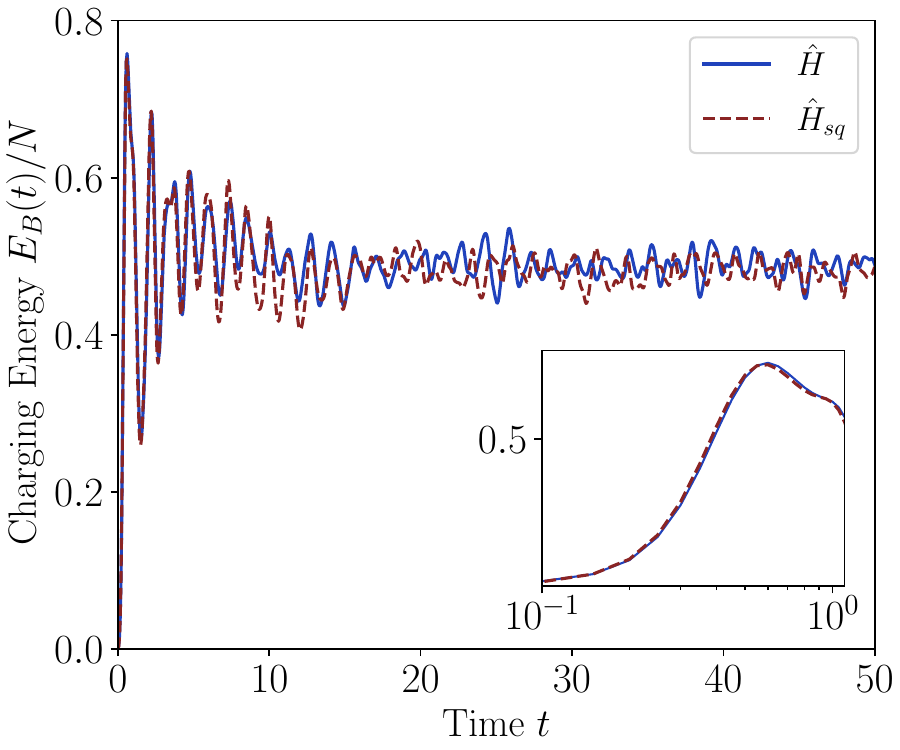}
    \caption{Dynamics in two mode quantum battery. Charging energy per spin as a function of charging time for the quantum battery with $N=100$ spins, $\Delta_c = \Delta_s = 1~\omega_c$, $\Delta_b = 101~\omega_c$, $g_c = 0.05~\omega_c$, $g_b = 2~\omega_c$ and $\delta_b=-1$. Solid lines correspond to the dynamics generated from full Hamiltonian $\hat{H}$ in Eq.~\eqref{Eq:Hlab} of the main text and dashed lines represent the effective dynamics generated from SW transformed Hamiltonian $\hat{H}_{sq}$ in Eq.~\eqref{Eq:HamTAT}. The inset zooms the region around first peak of energy.}
    \label{fig:A1}
\end{figure}

\section{Errors in Schrieffer-Wolff dynamics\label{AppA2}}
Eliminating the cavity mode $b$ via a Schrieffer–Wolff (SW) transformation {of the Hamiltonian $\hat{H}$ in Eq.~\eqref{Eq:Hlab}} inevitably introduces errors in the charging dynamics, which in general grow with the evolution time. Consequently, the effective SW Hamiltonian {$\hat{H}_{sq}$} in Eq.~\eqref{Eq:HamTAT}} is not expected to remain accurate at very long times. In this work, however, we restrict our analysis to the dynamics around the first energy peak, where the SW description remains controlled. The error in the dynamics at the first peak scales as $\mathcal{O}\left(g_b\sqrt{N}/\Delta_-\right)$~\cite{Gong2020a,Gong2020b}. For the worst-case parameters used in our simulations, $N=100$, $g_b=2.0~\omega_c$, $\Delta_-= 100~\omega_c$, yielding $g_b\sqrt{N}/\Delta_-= 0.2$, so that the effective description remains quantitatively reliable in the time window of interest. Figure~\ref{fig:A1} shows the charging energy dynamics using full model Hamiltonian {$\hat{H}$} and the SW transformed Hamiltonian {$\hat{H}_{sq}$}. It can be seen that both dynamics match at short times (around first peak), but the errors become significant at longer times.

\bibliography{references}

\end{document}